# Electrochemical reduction of thin graphene-oxide films in aqueous solutions: restoration of conductivity

Dalibor Karačić[1], Sanjin J. Gutić[2], Borislav Vasić[3], Vladimir M. Mirsky[4], Natalia V. Skorodumova[5], Slavko V. Mentus[1,6], Igor A. Pašti[1,5]*

[1] *University of Belgrade – Faculty of Physical Chemistry, Belgrade, Serbia*
[2] *University of Sarajevo, Faculty of Science, Department of Chemistry, Sarajevo, Bosnia and Herzegovina*
[3] *Institute of Physics Belgrade, University of Belgrade, Belgrade, Serbia*
[4] *Institute of Biotechnology, Department of Nanobiotechnology, Brandenburgische Technische Universität Cottbus-Senftenberg, Germany*
[5] *Department of Materials Science and Engineering, School of Industrial Engineering and Management, KTH – Royal Institute of Technology, Stockholm, Sweden*
[6] *Serbian Academy of Sciences and Arts, Belgrade, Serbia*

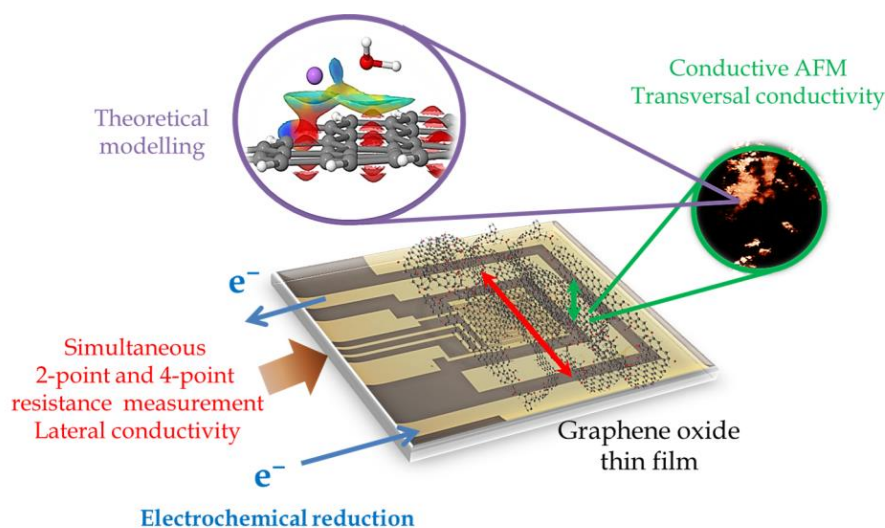

*corresponding author
Email: igor@ffh.bg.ac.rs

**Abstract**

Graphene oxide finds applications in different fields of science, including energy conversion. Electrochemical reduction of graphene oxide (GO) significantly improves its conductivity. However, the kinetics of this process depends on the solvent, supporting electrolyte, pH, and numerous other factors. Most studies report the macroscopic views and *ex-situ* properties of reduced GO. To expand the knowledge about GO reduction, in this study, we used cyclic voltammetry (CV), simultaneous 2 points and 4 points resistance measurement (s24), conductive atomic force microscopy (AFM), and theoretical calculations. Using CV, we demonstrated that the choice of supporting electrolyte (KCl or LiCl) influences the potential range in which electrochemical GO reduction occurs. The activation energy of this process was estimated to be below 30 kJ mol$^{-1}$ in both electrolytes, being significantly lower than that required for thermal reduction of GO. Simultaneous *in situ* s24 resistance measurements suggest that GO films reach a highly conductive state at deep negative potentials, with an abrupt, irreversible switch from non-conductive to the conductive state. However, conductive AFM presents a more exact picture of this process: the reduction of GO films starts locally while the formed conductive islands grow during the reduction. This mechanism was confirmed by theoretical calculations indicating that the reduction starts on isolated oxygen-functional groups over the GO basal plane, while clustered OH groups are more difficult to reduce. The presented results can help in tailoring reduced GO for a particular electrochemical application by precisely controlling the reduction degree and percentage of the conductive area of the reduced GO films.

# 1. Introduction

Since its discovery, graphene has found numerous applications in various fields of science and technology, including, but not limited to, energy conversion and storage, biomedical engineering, electronics, sensors, aerospace applications, and others [1]. A vast number of existing reports list some of the well-known properties of the idealized graphene sheet (pristine infinite 2D sheet of hexagonal carbon), such as large specific surface area, high carrier mobility, and good mechanical properties as a determinant for graphene applications [1]. However, this can be considered as an exaggeration as pristine graphene is practically non-existing. Moreover, even experimentally realized, ideal graphene would be practically useless in some technologies, like electrochemical energy conversion and storage [2]. This observation is clearly demonstrated by the fact that, in comparison to the basal plane, the graphene edges provide four orders of magnitude higher specific capacitance associated with double layer charging, much faster electron transfer rate, and fair electrocatalytic activity [3]. Thus, for most electrochemical devices, a material with some features of the perfect graphene (such as high surface area and high electrical conductivity) but with a significant number of structural imperfections and chemical moieties [4–6] is required.

Desired chemical moieties and defects in graphene can be introduced by using, for example, selective oxidation [7]. However, a more viable route is the production of reduced graphene oxide (rGO) from graphene oxide (GO) obtained by chemical exfoliation of oxidized graphite by chemical, thermal, or electrochemical reduction [8,9]. Upon GO reduction, irrespective of the particular reduction technique, enhanced capacitance and charge transfer properties of rGO are evidenced, along with the decreased O/C ratio connected to enhanced conductivity [8,10]. However, one has to find a proper balance between the amount of oxygen functional groups and the conductivity of rGO to reach optimal capacitance [11,12]. Such precise tuning and controllable modifications of structural and chemical properties of graphene oxide are of general importance for applying graphene-based materials in different electrochemical systems [7]. They can be easily achieved using electrochemical reduction, employing electrode potential to control the reduction degree of final rGO.

For electrochemical reduction, GO can be either suspended in an electrolyte solution or deposited directly on the electrode surface as a thin film [9]. The reduction was previously performed in aqueous and non-aqueous electrolytes [13], while characterization is typically performed using *ex-situ* techniques (X-ray photoelectron spectroscopy, XPS, Fourier Transform Infrared Spectroscopy, FTIR, Raman spectroscopy). It was shown [13–16] that different O-functional groups are reduced at different potentials, while the effects of solvent and pH on the behavior of resulting rGO were evidenced. The process is rather complex, being dependent on the nature of supporting electrolytes [12]. The effects of the GO reduction on its electrochemical properties are clear. However, electrochemically produced rGO films are typically characterized using various techniques that provide a macroscopic average of the final rGO. For example, XPS, FTIR, and Temperature Programmed Desorption (TPD) revealed the presence of different functional groups on the rGO surface, while Raman spectroscopy was routinely employed to investigate a structural disorder in rGO. In addition, macroscopically averaged properties are obtained using electrochemical tests with the ferro/ferricyanide system to assess charge transfer properties of rGO. Thus, we are still not fully aware of how electrochemical reduction of GO proceeds at the submicron scale in terms of local distribution of reduced domains and their impact on the conductivity of electrochemically reduced GO films. At the same time, the conductivity of the electrode material affects its overall electrochemical performance. Adjusting conductivity of rGO is crucially related to the optimization of capacitive response [12], but also the understanding of the GO reduction process is necessary for the efficient formation of rGO-based composites with high catalytic activity [17]. The lack of atomic-level information is clear, and during the writing of this paper, a study combining scanning transmission X-ray microscopy and Kelvin probe force microscopy [18] was published. However, in this particular work, Rodriguez *et al*. focused on the local work function variations, showing that they arise due to the presence of oxygen functional groups, being of crucial importance for photovoltaic behavior [18].

In this contribution, we address the electrochemical reduction of thin films of GO in aqueous solutions of LiCl and KCl using, for experimental considerations, cyclic voltammetry, *in situ* simultaneous 2-point and 4-point resistance measurements, and *ex-situ* conductive atomic force microscopy. The results show that the electrochemical reduction of GO has low apparent activation energies while the reduction process commences locally. The conductive islands start to coalesce at deep negative potentials making the film completely conductive. Furthermore, the differences in the GO film reduction process in LiCl and KCl solutions are explained using semiempirical quantum chemical and Density Functional Theory calculations.

## 2. Experimental

*2.1. Electrochemical reduction of GO thin films*

The reduction of GO thin films at different temperatures and in different electrolytes was done as follows. First, aqueous GO suspension (standard solution, 4 mg ml$^{-1}$; Graphenea, Spain [19]) was diluted to obtain 1 mg ml$^{-1}$ in a 6:4 water/ethanol (v/v) mixture. After sonication, 10 µl of the obtained GO suspension was drop-casted onto the copper foil and dried under vacuum at room temperature. Pt foil served as the counter and Ag/AgCl (saturated KCl) as the reference electrode. All potentials are indicated *versus* this reference electrode. Ivium Vertex.One potentiostat was used for the measurements. Electrochemical measurements were performed in 0.1 mol dm$^{-3}$ aqueous LiCl and KCl solution at pH adjusted to ~ 6.6±0.1. GO reduction was done by a single potentiodynamic scan at 10 mV s$^{-1}$ between −0.5 (starting potential) and −1.6 V (KCl solution) or −1.8 V (LiCl solution). The measurements were done at four temperatures: 7.5, 20, 30, and 40 °C. For this purpose, the cell containing the working and counter electrodes was thermostated while the compartment with the reference electrode was held at 25 °C and connected to the cell using a salt bridge with a Lugin capillary. The *iR* drop was compensated using the positive feedback scheme, the value of the electrolyte resistance was measured using impedance measurement at 100 kHz at −0.5 V *vs.* Ag/AgCl.

*2.2. In situ resistance measurements during the GO thin film reduction*

The electrical resistance of (r)GO films was measured as described elsewhere [20,21] using simultaneous two- and four-point techniques (s24-technique). Shortly, 50 mV pulses of alternating polarity were applied, and the total resistance was calculated. This resistance includes both the bulk resistance of the sensing composite and the contact resistance ($R_c$). It corresponds to the resistance that would be measured using the conventional two-point technique, and it is denoted here as $R_2$. Simultaneously, the potential drop between the inner electrodes of the interdigitated structure (see further) was measured with the electrometric voltmeter (Keithley-617) and used to calculate the bulk material resistance. This resistance corresponds to that measured by the four-point technique, and we denote it here as $R4$. Then the contact resistance can be determined as $R_c = R2 - \alpha \cdot R4$. For the case of the interdigitated gold electrode used in this work, the parameter $\alpha$, which primarily depends on the geometry, is determined to be around 3 [22]. Alternatively, one can use the $R2/R4$ ratio to estimate the contribution of the contact resistance into $R_2$. For the geometry of our sensors, the $R2/R4$ ratio should be around 3 if no contribution of the contact resistance into $R2$ occurs [20,22]. The electrodes were prepared by drop-casting 2.5 µL of diluted GO dispersion (0.04 wt.% or 0.004 wt.%, Graphenea, Spain [19]. Upon drying the GO film, the electrodes were transferred into the electrochemical cell, and the measurements were done as described below.

The measurement system has a sampling interval of around three seconds which imposes the lower limit for the measurements under potentiodynamic conditions. For this reason, the reduction of the thin film of GO was made in steps using a program consisting of cyclic voltammetry steps (one cycle to the given cathodic vertex potential; potential sweep rate of 10 mV s$^{-1}$) and potentiostatic steps for the resistance measurements. Each scan started from −0.5 V *vs.* Ag/AgCl (common anodic to vertex) and was performed to the cathodic vertex starting from −0.8 V down to −1.3 V *vs.* Ag/AgCl (KCl solution) or −1.4 V *vs.* Ag/AgCl (LiCl solution). The cathodic vertex was reduced by 0.1 V every cycle, from −0.8 V to −1.3 V (KCl solution) or −1.4 V (LiCl solution). After each cycle to a given cathodic vertex potential, s24 measurements were performed under potentiostatic conditions (−0.5 V *vs.* Ag/AgCl), and the resistance values were collected for 1 minute. The measured values were averages over this sampling period. The outermost contacts of the interdigitated gold electrode sensor were connected to the AutoLab potentiostat PGSTAT12.

*2.3. SEM, Raman spectroscopy and AFM characterization of reduced GO films*

The GO thin films reduced during the *in-situ* resistance measurements were characterized using Raman spectroscopy. Raman spectra (excitation wavelength 532 nm) were collected on a DXR Raman microscope (Thermo Scientific, USA) equipped with an Olympus optical microscope and a CCD detector. The laser beam was focused on the sample using objective magnification 50×. The scattered light was analyzed by the spectrograph with a 900 lines mm$^{-1}$ grating. Laser power on the sample was kept at 2.0 mW.

SEM with EDX was done using Phenom ProX (Phenom, The Netherlands). Chemical analysis was done using an acceleration voltage of 15 kV. Presented chemical compositions are averaged over five sampling points, with a sampling time of 1 min each.

An analogous series of the reduced GO thin films were produced for the analysis by atomic force microscopy (AFM). The reduced GO films were prepared on copper foils and reduced in KCl or LiCl solutions (0.1 mol dm$^{-3}$) at potentials between −0.8 and −1.5 V using potentiostatic steps of 10 s duration. The reduction was made with a 0.1 V resolution. After electrochemical reduction, the electrode with GO film was dried in ambient conditions. AFM-based analysis of graphene-oxide films was done using Tegra Prima. Local current maps were measured by conductive AFM (C-AFM), and diamond coated probes DCP20 from NT-MDT. During C-AFM measurements, the bias voltage of 10 V was applied on the backside of graphene-oxide films, while the grounded probe scanned the top surface in contact AFM mode. Therefore, the measured current corresponds to the vertical charge transport through the films. In order to provide a stable AFM tip-sample contact during imaging, the normal load was used as a control parameter. For each reduction potential, five 10×10 µm² current maps were measured on different sample locations. In order to characterize the sample conductivity, two characteristic values were calculated for each current map: the average current and the ratio between the conductive and scanned 10×10 µm² area (in percents). The resolution of AFM images was 256×256 data points *per* image, so the average current was calculated as a mean value of the corresponding two-dimensional matrix. In order to calculate the ratio between the conductive and scanned area, $I_{thr}$=0.1 nA was selected as a current threshold, meaning that points on a sample surface were counted as conductive ones only if the measured current was higher than $I_{thr}$. Finally, the average current and conductive area for each reduction potential were obtained by averaging results obtained on five different sample locations.

*2.4. Theoretical calculation*

Theoretical calculations were performed to understand better the interactions of metal atoms/cations with GO surface and functional groups. The GO model was set as a finite sheet with H-saturated edges containing several oxygen functional groups. We have investigated the interactions of metal atoms with clustered and isolated OH groups on this GO sheet model. The details on the calculation setup are provided below.

The first-principle DFT calculations were performed using the Vienna *ab initio* simulation code (VASP) [23–25]. The Generalized Gradient Approximation (GGA) in the parametrization by Perdew, Burk, and Ernzerhof [26] combined with the projector augmented wave (PAW) method was used [27]. Cut-off energy of 450 eV and Gaussian smearing with a width of $\sigma$ = 0.025 eV for the occupation of the electronic levels were used. Brillouin zone was sampled using $\Gamma$-point only. During structural optimization, the relaxation of all atoms in the simulation cell was allowed. The relaxation proceeded until the Hellmann–Feynman forces acting on all the atoms became smaller than $10^{-2}$ eV Å$^{-1}$. Spin polarization was included in all calculations.

Semiempirical calculations were done using MOPAC2016 code [28] with PM7 method [29]. Full structural relaxation was done. The analysis was performed in the presence of water as a solvent. The solvent was included in the analysis implicitly, using the Conductor-like Screening Model (COSMO) method [30]. Additionally, 1-3 explicit water molecules were introduced into the solvation spheres of Li or K.

Visualization was done using VESTA [31] and Jmol [32].

## 3. Results and Discussion
### 3.1. Electrochemical reduction of GO films – overview and activation energies

First, electrochemical reduction of thin GO films was investigated in 0.1 mol dm$^{-3}$ LiCl and KCl solutions using cyclic voltammetry. As previously demonstrated [12] the reduction process is fast and irreversible, where one major reduction peak is observed (**Figure 1**). The reduction commences at lower potentials in LiCl solution, and with decreasing the temperature from 40 to 7.5 °C, the reduction peak shifts to lower potentials. The same situation is observed for KCl, but the separation of reduction peaks increases upon decreasing the temperature, while the reduction peaks get broader and lower in the current (absolute values).

The temperature dependence of measured current can be expressed in a general form as:

$$i(E) = C(E) \times \exp\left(-\frac{E_{act}(E)}{RT}\right) \qquad (1)$$

where $C$ assembles all the constants (including electrode potential dependent term), and $R$ and $T$ are universal gas constant and absolute temperature while ($E_{act}(E)$) is the apparent activation energy for this process. Therefore, for a given electrode potential $E$,

$$\ln i(E) = \ln C(E) - \frac{E_{act}(E)}{RT} \qquad (2)$$

assuming that $C(E)$ does not depend on temperature. Upon processing recorded $I$-$E$ curves, $E_{act}$ was found to increase as the cathodic polarization increases up to the potentials corresponding to approx. 70 % of the cathodic peak recorded at 40 °C, upon which it starts to decrease. The values are slightly lower in KCl (22 kJ mol$^{-1}$) than in LiCl (27 kJ mol$^{-1}$).

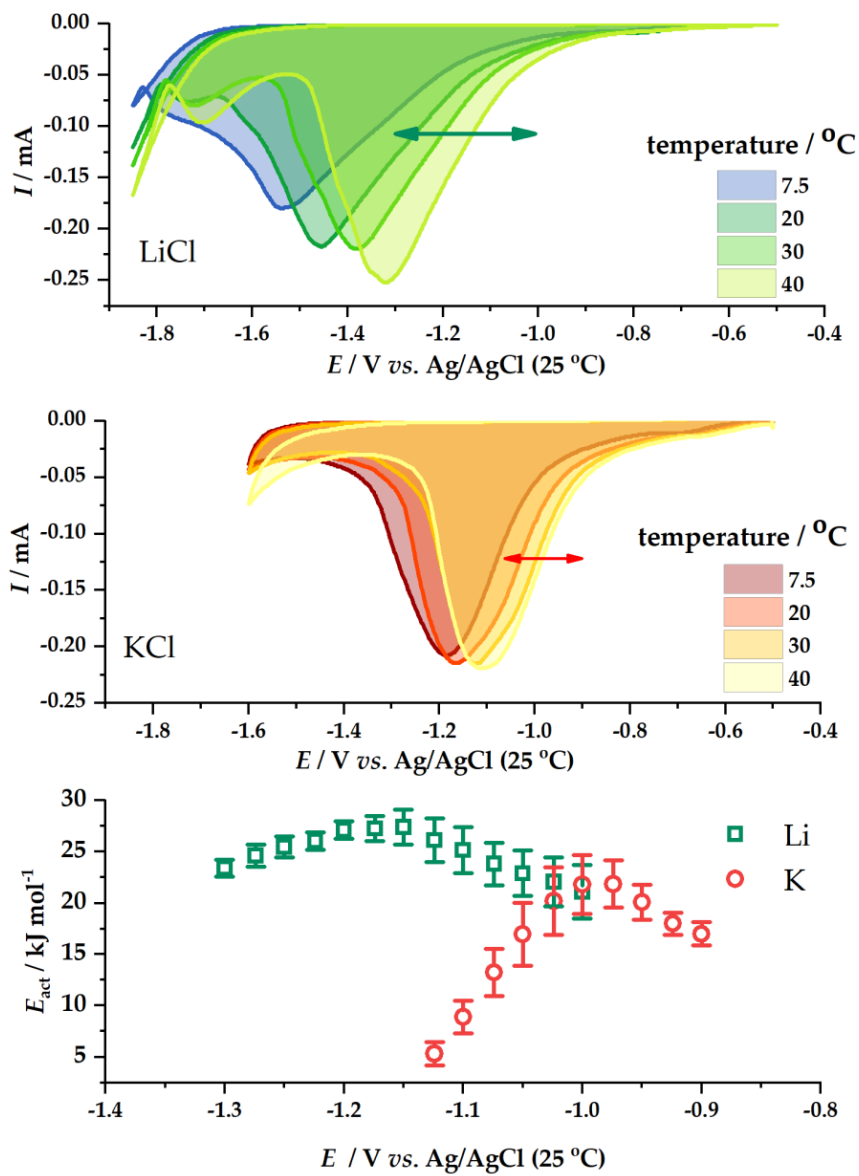

**Figure 1. Reduction of GO thin films as seen using cyclic voltammetry**: GO thin film reduction in 0.1 mol dm$^{-3}$ LiCl (top) and KCl (middle) at four different temperatures, potential sweep rate: 10 mV s$^{-1}$. The arrows indicate the potential ranges in which the apparent activation energies for GO reduction were calculated.

The values of $E_{act}$ obtained near room temperature are expected to be significantly lower than those at elevated temperatures. Experimental data confirmed it. For example, the resistivity measurements on individual single-layer GO platelets at temperatures above 140 °C give the activation energy of (155±4) kcal mol$^{-1}$. On the other hand, the TPD measurements of multilayer films of GO platelets give the activation energy of (134±17) kJ mol$^{-1}$ [33]. During the low-temperature annealing, the resistivity measurements gave the activation energy of 1.65 eV (i.e., 159 kJ mol$^{-1}$) [34] it was ascribed to the processes of desorption of epoxy and alkoxy oxygen atoms together with carbon [35] and to the restoration of non-oxidized graphene domains. Obviously, the electrochemical reduction of GO being completed within minutes is a much faster process than the low-temperature annealing taking 2-4 h [33]. Finally, we must comment that the values of activation energies derived here and in other papers should be interpreted carefully. This unit should be understood as the energy *per* Avogadro number of chemical bonds. Namely, due to a scatter in the content of surface oxygen, one mole of GO is poorly defined, and molar values cannot be undoubtedly associated with converting one mole of GO to rGO. These results are better to be interpreted at one mole of chemical bonds present in GO which get broken during the reduction process and averaged over the ensemble of groups present in a particular GO sample. However, we note that the change of the nature of dominant oxygen functional groups in a given GO sample cannot compensate for the difference between $E_{act}$ for electrochemical reduction against low-temperature annealing, since, in the latter case, the process is much slower.

*3.2. Lateral conductivity and reduction of films of GO.*

The resistance of GO films during annealing is decreasing for seven orders of magnitudes [34], as a consequence of thermally induced oxygen release. We have investigated the changes of resistance using simultaneous 2-point and 4-point measurements, which provide information on the lateral and contact resistances of GO films during the electrochemical reduction (**Figure 2**). The results were found to depend on the film thickness. For thinner films, $R$2 was found to decrease slowly from very high values (10$^7$ Ω) with increasing cathodic potential, until −1.1 V is reached. Then, $R$2 decreases by two orders of magnitude for the film reduced in KCl and continues to decrease for the film reduced in LiCl (**Figure 2**, c). The $R$4 values show a similar trend but with a characteristic increase (**Figure 2**, c) at potentials close to the reduction peak potential (**Figure 1**).

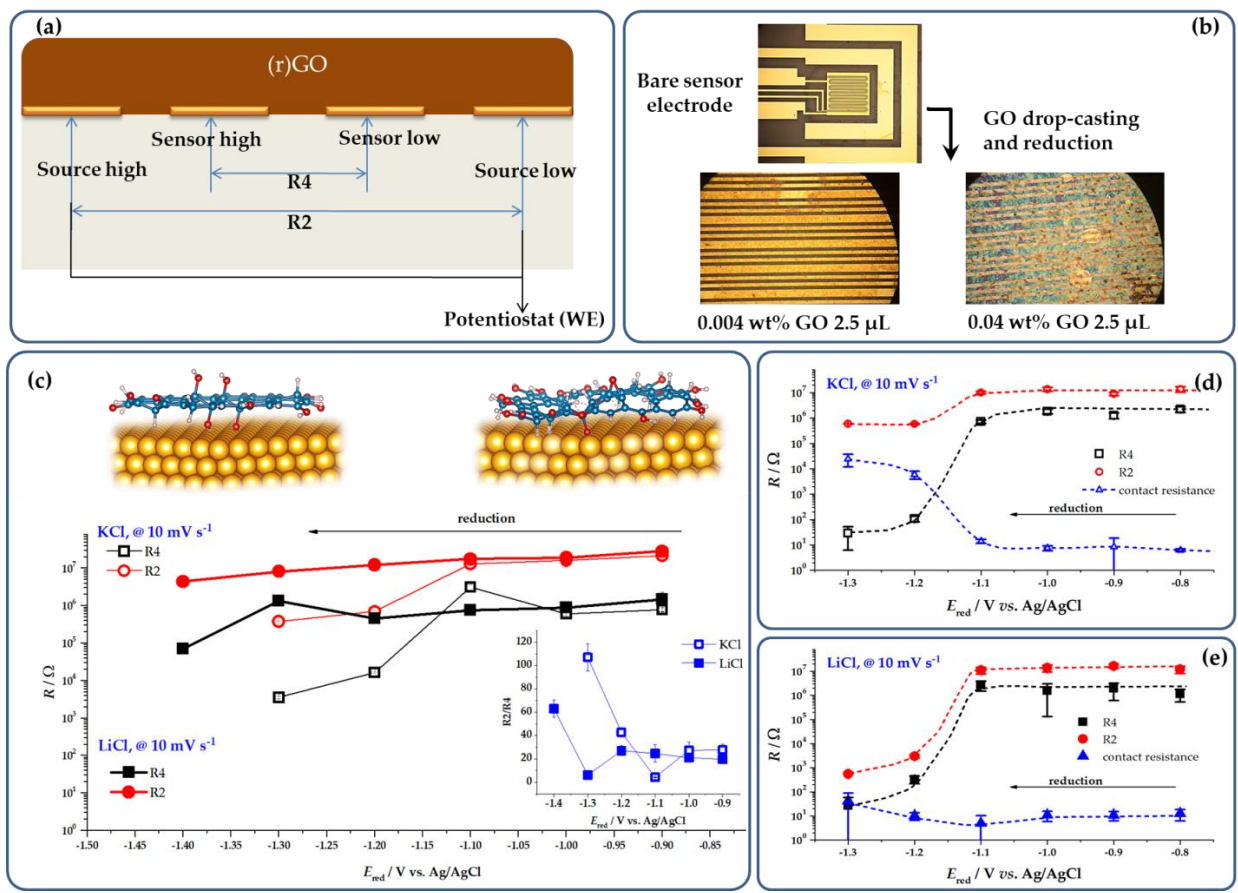

**Figure 2. Reduction of GO deposited on the gold interdigitated electrode as seen using s24 measurements**. (a) The connection of the interdigitated gold sensor electrode to the resistance measurement system and the potentiostat, components for s24 measurements are not shown, (b) optical micrographs of the interdigitated sensor electrode and magnified parts of the electrode upon reduction of the GO film of different thickness, (c) s24 measurements for GO film obtained upon drop-casting of 2.5 µL of 0.004 wt.% GO dispersion in KCl and LiCl at different reduction potential corresponding to non-reduced GO and reduced GO film (schematically shown for two vertex potentials), inset gives the $R2/R4$ ratios for two electrolytes, (d) s24 measurements for GO film obtained upon drop-casting of 2.5 µL of 0.04 wt.% GO dispersion in KCl, with evaluated contact resistance, (e) the same as for (d) but for the reduction in LiCl solution.

Looking at the $R2/R4$ ratios (**Figure 2**, c, inset), for the non-reduced GO films, the ratio is much higher than the theoretical value for zero contact resistance (i.e., ~ 3), amounting to ~20. These values suggest a poor contact between the GO film and the gold electrode. When the reduction process starts, the value of the $R2/R4$ ratio drops down to the value of ~3, after which they increase significantly, with the increment being more pronounced for KCl. The behavior of thicker films is similar (**Figure 2**, d and e), but the resistance changes are much more pronounced, and $R4$ values change for six orders of magnitude for the films reduced in both LiCl and KCl. However, in the KCl solution, the $R2$ drops by only one order of magnitude, and the contact resistance reaches the values of $10^4$ Ω. For the films reduced in LiCl, $R4$ changes for four orders of magnitude, and the contact resistance reaches the values of 60 Ω.

The large differences in the behavior of the contact resistance between two electrolytes can be explained by faster reduction of GO in KCl, resulting in pronounced evolution of CO, $CO_2$, and $H_2O$ from the GO sheets [36]. However, one should also consider the $H_2$ evolution on the gold electrode, weakening the contact between the GO film and the electrode. Namely, the $H_2$ evolution on the gold electrode is more pronounced in $K^+$-containing electrolyte than in $Li^+$-containing one [37]. Thus, more intense $H_2$ evolution in KCl can also contribute to the weaker contact between the gold electrode and the reduced GO film. It is also interesting to observe a certain improvement in the contact before reduction. The initially evolved CO, $CO_2$, and $H_2O$ probably improve the compactness of the film and the contact with the Au electrode, as out-of-plane surface groups start to be removed and the stacking of the layers is improved. Such compactization could also be associated with previously observed changes (using *in situ* surface-enhanced infrared spectroscopy) in the double layer at the GO film–electrolyte interface and hydrogen bonding of intercalated water between the GO sheets at moderate reduction potentials before reducing O-functional groups commences [38]. However, as the evolution of gasses continues, the formed microbubbles weaken the contact between the reduced GO sheet and the Au electrode. Finally, the non-conductive to conductive state transitions are clear for thinner films, seen by measured $R2$ and $R4$ in LiCl and KCl solutions (**Figure 2**, c). However, for thicker films, the change of resistance appears at the same potential (just below −1.1 V, **Figure 2**, d and e). Such behavior can be understood assuming that for thicker films, a sufficient amount of reduced GO sheets is formed in both electrolytes so that the conductive network is built and the lateral conductivity of the films increases. Another possibility is that upon repeated cycling of a GO film to progressively higher cathodic vertex potentials, the degree of reduction increases over the value corresponding to the degree reached during a single scan to a given cathodic vertex. This scenario is feasible as we found that electrochemical GO reduction is also sensitive to the scan rate and shows typical irreversible character so that higher degrees of reduction can be achieved at lower potentials with very slow potential scans.

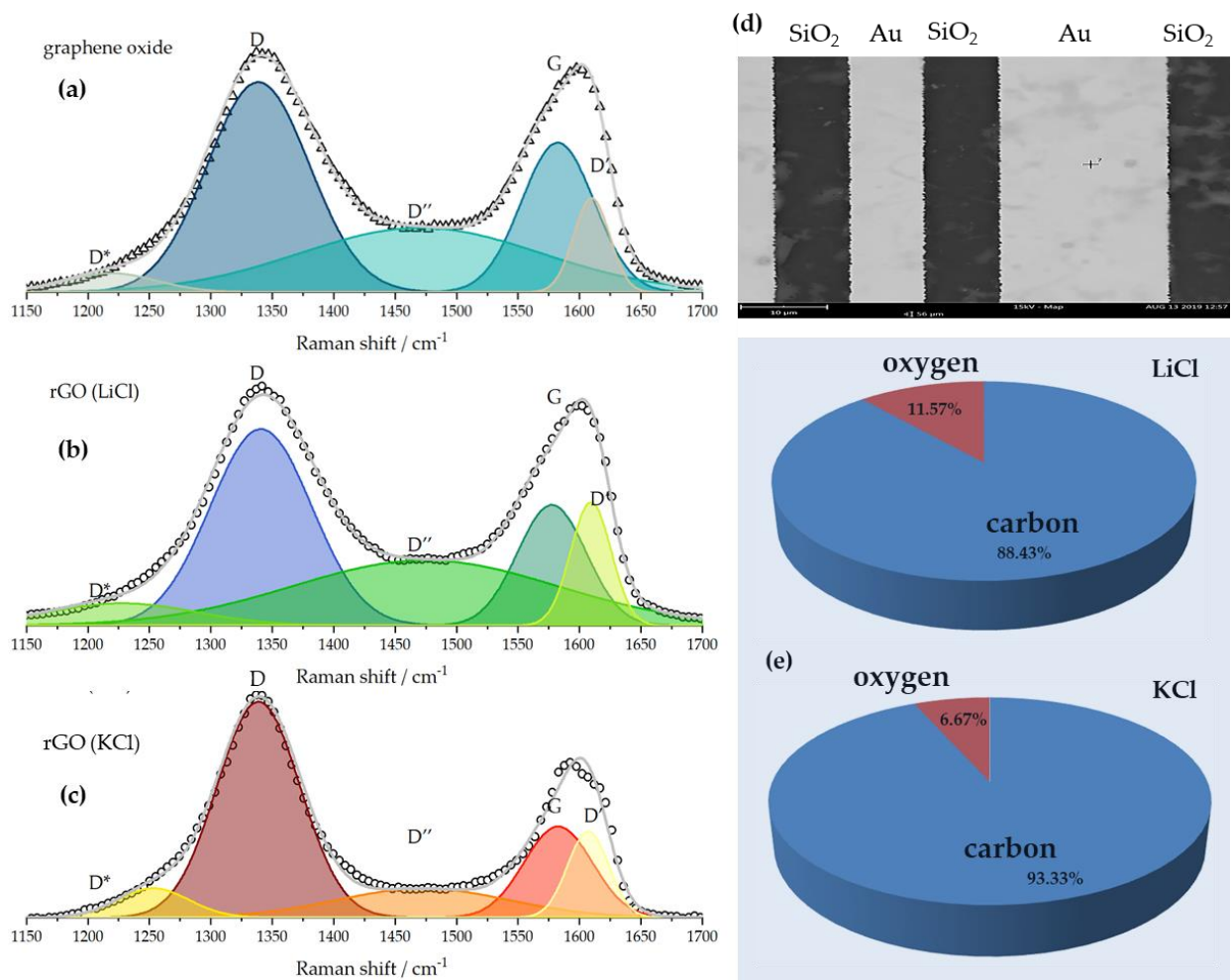

**Figure 3.** *Ex-situ* **characterization of reduced GO films on the gold interdigitated electrode.** Raman spectra of GO thin film deposited on gold interdigitated sensor electrode (2.5 µL of 0.004 wt.% GO) before reduction (a) and after electrochemical reduction in LiCl solution (b) and KCl solution (c). Both films are reduced using a single cyclic voltammetry scan from −0.5 to −1.3 V vs. Ag/AgCl with a sweep rate of 10 mV s$^{-1}$. On the right SEM image of the interdigitated gold electrode with rGO film is given (d), and the results of EDX chemical analysis of the GO film in LiCl (C:O ratio 7.6±1.2) and KCl (C:O ratio 13.9±3.4).

The reduced GO films on Au electrodes used in the resistance measurements were subjected to Raman and EDX analysis (**Figure 3**). The Raman spectra in the 1150-1700 cm$^{-1}$ region were deconvoluted into five components using Gaussian profiles. For this reason, the reported band intensity ratios are obtained using the peaks areas [39]. For non-reduced GO film, the G band is located at 1582 cm$^{-1}$, while the $I_D/I_G$ ratio is 1.97. For the film reduced in LiCl solution, the G band is at 1577 cm$^{-1}$, and for the GO film reduced in KCl, it is found at 1583 cm$^{-1}$. The corresponding $I_D/I_G$ ratios are 2.40 and 2.89, respectively. Considering a high oxygen content and practically no sp$^2$ bonded carbon in the used GO sample [19], the as-received GO sample can be located in stage 3 of the amorphization trajectory [39]. However, there is a clear D band compared to the tetrahedral amorphous carbon where the D band intensity is null. However, for fluorinated graphene with a high degree of sp$^3$-type defects, the D band intensity was also found not to be zero [40]. The C:O ratios found in reduced GO films (**Figure 3**, e) suggest that the film reduced in LiCl is reduced to a lower level than that reduced in KCl. As the G band is located at a slightly lower wavenumber and the $I_D/I_G$ ratio is smaller for the GO film reduced in LiCl, it can be placed in stage 2 of amorphization trajectory (positioned between nanocrystalline graphite and amorphous carbon-based on Raman spectrum) [39,41]. Formally, the film reduced in KCl can also be placed in the same stage but closer to stage 1, as indicated by a higher position of the G band and a higher $I_D/I_G$ ratio. This conclusion is also in line with the $I_{D'}/I_G$ ratios for reduced GO films, amounting to 0.60 and 0.57 for films reduced in LiCl and KCl solution, respectively [41].

The question is, which types of defects are present in the reduced films? Using the $I_D/I_{D'}$ ratios, it is, in principle, possible to resolve the type of defects present in graphene [41]. The non-reduced GO film has the $I_D/I_{D'}$ ratio of 6.3, close to the typical value for vacancy-like defects ($I_D/I_{D'}$ = 7, [41]). However, the same ratios for reduced GO films are smaller than for non-reduced GO, 3.99 and 5.07 for films reduced in LiCl and KCl, respectively. At least for the film reduced in LiCl, this might indicate the dominant presence of boundary-like defects. However, there is also a possibility that a hysteresis appears during the reduction of GO [39], as, in fact, reduction of GO goes along the ordering trajectory.

For this reason, it is rather difficult to derive a precise conclusion of the type of defects in reduced GO films. However, as CO and $CO_2$ evolve and the process is rather fast (tenths seconds *vs*. hundreds of minutes in thermal reduction procedure), it can be expected that reduced GO is rich in vacancies. This conclusion is aligned with the fact that the starting GO is also exceptionally rich in very large vacancies [19]. This is in line with previous conclusions that electrochemical reduction of GO cannot heal the vacancies initially present in GO [9]. In contrast, it was recently suggested [42] that electrochemical reduction favors the formation of $sp^3$-like defects over vacancies. However, we suspect that the reduction of GO films was incomplete in these experiments as performed only down to –1 V *vs*. Ag/AgCl electrode in NaCl solutions. This potential actually corresponds to very low reduction degrees of GO (**Figure 1**, and Ref. [12]).

*3.3. Reduction and transversal conductivity of electrochemically reduced GO films*

The reduced GO films were also analyzed using C-AFM. In contrast to s24 measurements, this method measured transversal conductivity through GO film. Typical current maps for several reduction potentials and the LiCl electrolyte are depicted in **Figure 4** (a). For the low reduction potential (−0.8 V), graphene-oxide films behave as insulators with a current below 120 pA (represented by a dark contrast) across the whole sample surface. Small bright patches appear in the current maps by an increase of the reduction potential till −0.9 V and −1.1 V leads to the appearance of small bright patches indicating a formation of conductive and spatially separated islands. At higher reduction potentials (−1.3 V), the current maps become inverted, consisting of a bright (conductive) surface with small and isolated dark (insulating) domains. At the highest reduction potential (-1.5 V), the dark domains practically disappear. Therefore, graphene-oxide films become highly conductive, containing only small insulating patches.

The changes in the current maps as a function of the reduction potential were quantified by calculation of the average current and relative area of conductive domains in the scanned areas. They are presented in **Figure 4**, (b) and (c), respectively, for both LiCl and KCl electrolytes. As can be seen, the reduction of graphene-oxide films has the same trend in both electrolytes, but the reduction in KCl starts earlier. At a reduction potential of −1 V, the average current is so high as ~9 nA, and the relative area of conductive domains is ~85%. In LiCl, a pronounced reduction happens later. Only at the reduction potential of −1.3 V almost the entire GO films become conductive while the average current jumps to 15 nA.

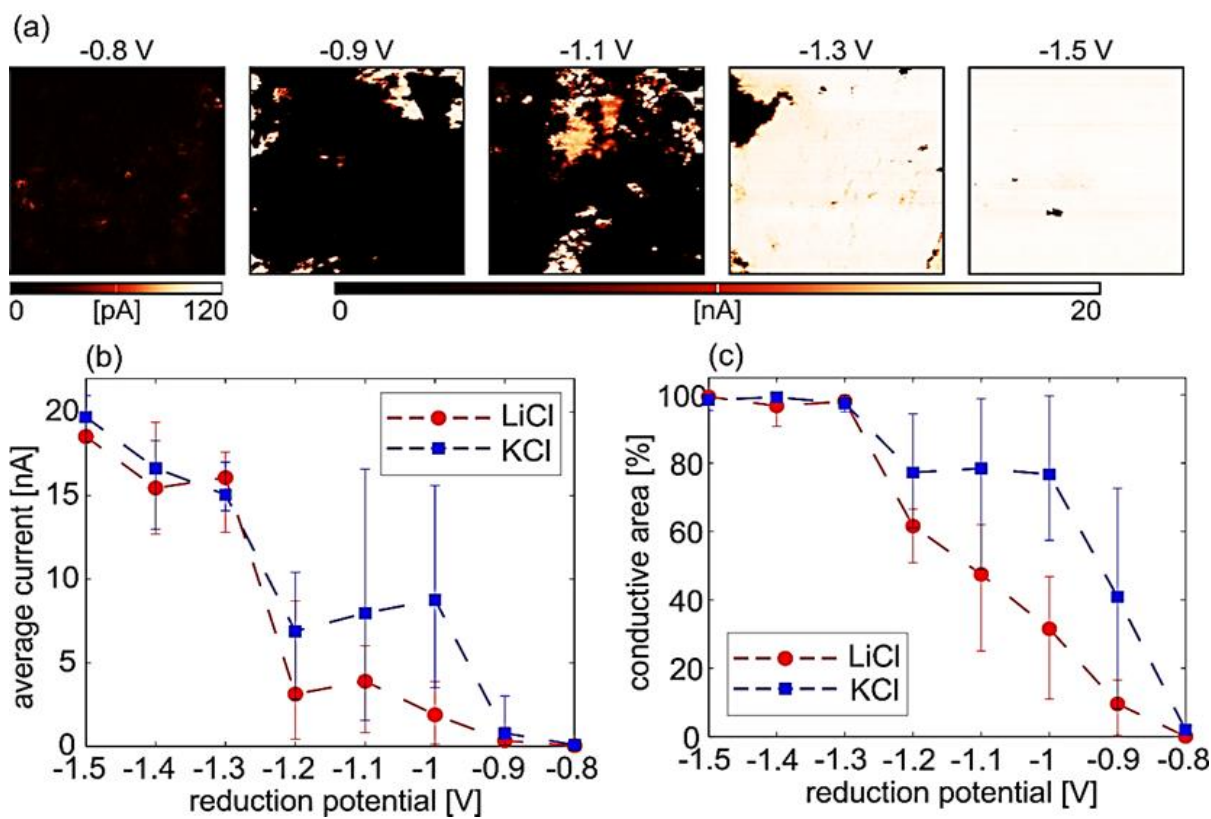

**Figure 4. The reduction of GO film as seen by C-AFM**. (a) Characteristic current maps for five selected reduction potentials measured in LiCl electrolyte. The scan size is 10×10 μm². The current scale is 20 nA except for the left map (-0.8 V), where the scale is so low as 120 pA because of the very low conductivity of unreduced samples. (b) Average current and (c) relative conductive area (calculated as a ratio between conductive and scanned, 10×10 μm² area, in percents) averaged on five different sample locations for each reduction potential.

Using the results of the cAFM measurements, one can conclude that the sharp decrease in the film resistance ($R4$) obtained using s24 measurements corresponds to the point when the conduction islands of reduced GO coalesce, forming a conductive network that bridges the gap between the electrodes in the 4-electrode sensor. We note that in the s24 measurements, the reduction of the GO film starts locally at the points where the film is in contact with the Au layer and spreads into the gap between two Au stripes (**Figure 3**, d). The distance between the Au stripes is under 10 μm, which matches the maximum lateral size of GO sheets used in this work. Hence, the measured resistance is also affected by the migration rate of reduced GO front in the gap region between two Au stripes. For this reason, the measured activation energy would be affected by the rate of the reduced GO front migration into the inter-Au region.

However, we note that this setup can also be used to estimate the rate of this migration. When a cyclic voltammetry sweep is performed at a higher rate (25 mV s$^{-1}$), we did not observe any resistance change, although the color change from yellow to black was visible. This observation indicates that the GO film was reduced at the places where it is in direct contact with the Au electrodes but not in-between them. As a cyclic voltammetry scan between −0.5 and −1.3 V at 25 mV s$^{-1}$ requires 32 seconds to complete, and the reduced GO film should cross 5 µm to connect into a conductive layer, it can be concluded that the reduced GO film progresses laterally at a rate below $1.6 \times 10^{-7}$ m s$^{-1}$ when not in direct contact with the current collector. However, this value is just a rough estimate as the reduction process does not occur during the entire potentiodynamic scan, while the reduced fragments of the GO film could act as current collectors for the reduction of adjacent GO parts.

*3.4. Theoretical modelling of GO reduction*

To model the process of GO reduction, we have performed a series of semiempirical and DFT calculations on a model of GO sheet (**Figure 5** and **Figure 6**). According to the Lerf–Klinowski model [43], epoxide and hydroxyl groups dominate the GO basal plane [44,45]. As we have previously discussed the differences in alkali and alkali earth metals with epoxide [12], here we focus on hydroxyl groups. This is also important considering that OH groups were recently found to be the first to reduce electrochemically [18]. The model contained several OH groups clustered together, knowing that such type of sp$^3$ defects appears in dimers or clusters [46,47]. The formal stoichiometry of the considered mode was $C_{58}H_{33}O_6$. An additional model was constructed so that it also contained one isolated OH group over the GO basal plane beside a cluster of OH groups (stoichiometry $C_{58}H_{34}O_7$).

In semiempirical calculations, water was added implicitly as a solvent, but we also considered explicit solvation of Li$^+$ and K$^+$ ions in our models. For Li$^+$, we have considered up to three explicit H$_2$O molecules in the hydration sphere, while for K$^+$ we considered systems with one or no H$_2$O molecules. When considered metal ions were brought in contact with the OH groups cluster, no detachment of OH groups and formation of MOH unit was observed. The same was when considered the interaction of metal ions with an isolated OH group. However, the oxygen atom in the OH groups is susceptible to electrophile attack. Thus, we added the H atom close to the O in the OH group interacting with the metal cation in the next step. The addition of the H atom is analogous to the addition of H⁺ through the electrolyte and one electron through the external electric circuit during the electrochemical reduction of GO, as we considered the whole systems as singlets and bearing one positive charge. A spontaneous detachment of the metal-H₂O complex is observed upon the relaxation in all the cases (**Figure 5**).

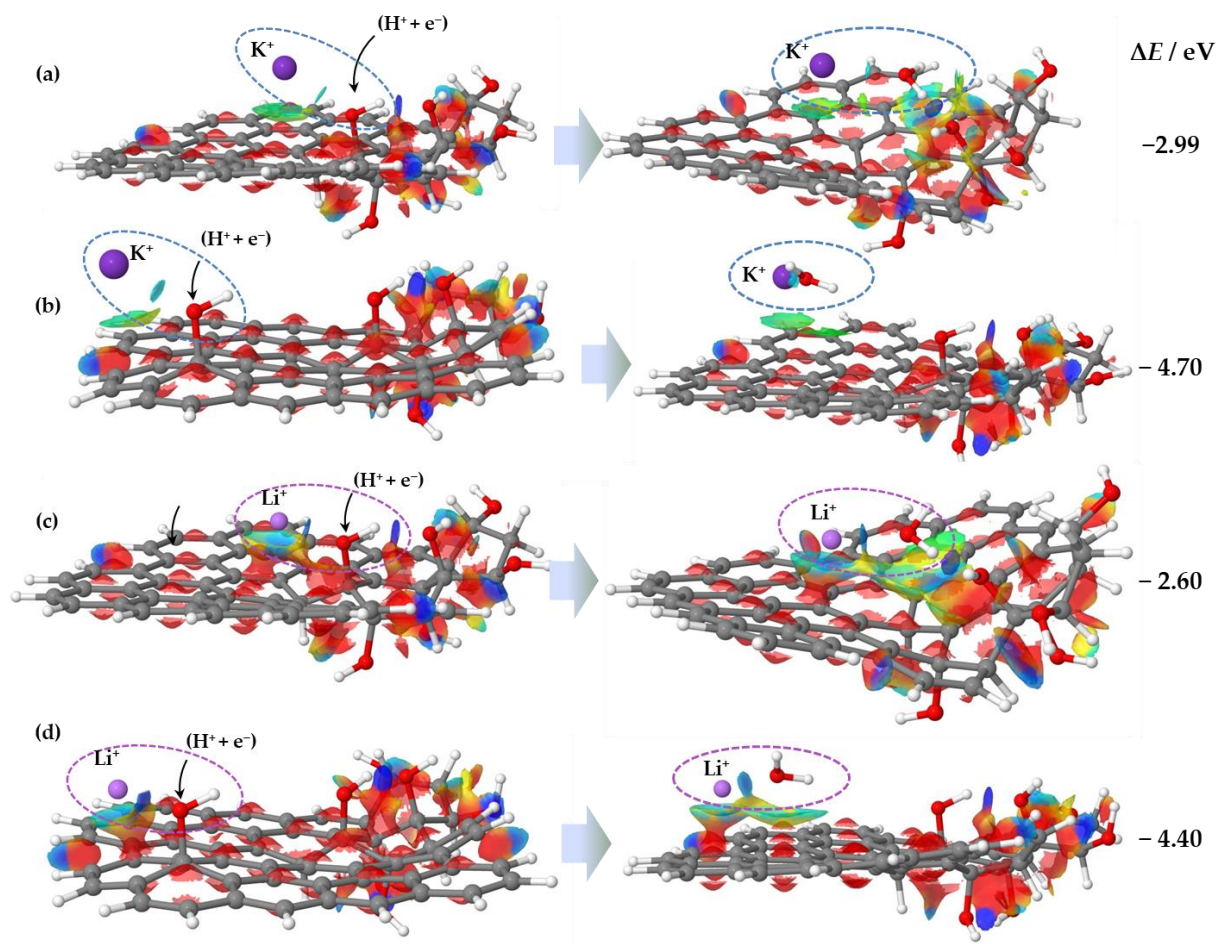

**Figure 5. GO reduction as seen by semiempirical quantum chemical calculations**. (a) Optimized structure of (K⁺ + GO) before and after the addition of one hydrogen (H⁺ + e⁻) to the system when K⁺ interacts with agglomerated OH groups, (b) optimized structure of (K⁺ + GO) before and after the addition of one hydrogen (H⁺ + e⁻) to the system when K+ interacts with an isolated OH group at the GO sheet, (c) the same as for (a) but for the interaction with Li⁺, (d) the same as for (b) but for the case of Li⁺. At the right, energy balances are provided. All the systems are calculated as singlets having one positive charge. Isosurfaces show the regions of non-covalent interactions present in the studied systems.

The main results that should be emphasized here are the following: (i) the total energy balance is always more negative for K⁺ compared to the analogous scenario with Li⁺ ions, and (ii) it is always energetically more favorable to remove an isolated OH group than the one from the cluster (**Figure 5**, see energy balances). The first result can be understood by extensive solvation of Li⁺ ion, which screens the Li⁺ ion and weakens the interaction with the detached H₂O molecule. The second result can be understood by stabilizing OH clusters due to the extended sp³ hybridization and the non-covalent interactions that stabilize an OH cluster (**Figure 5**, isosurfaces).

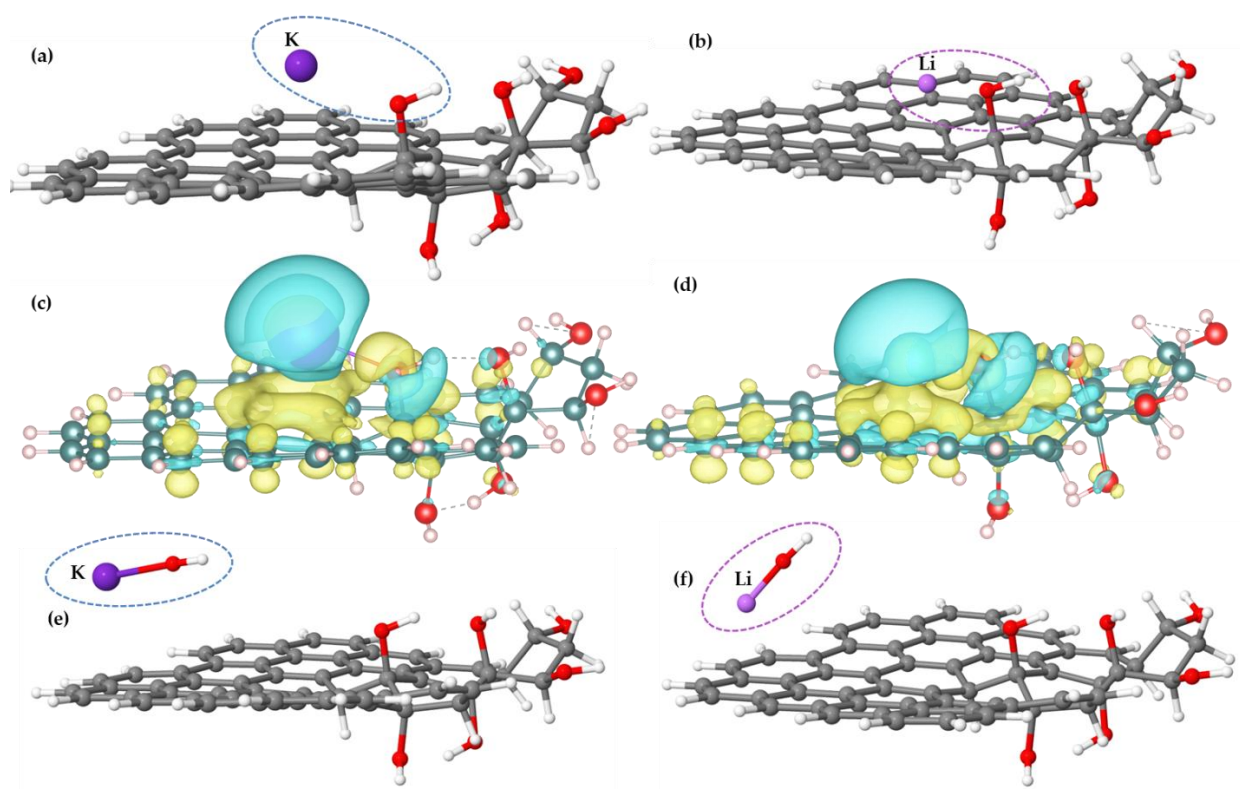

**Figure 6. GO reduction as seen by DFT**. Optimized structures for the interaction between K atom (a) and Li atom (b) with agglomerated OH groups, and the corresponding charge different plots (K – (c), Li – (d); isosurface values 1.2×10⁻³ e Å⁻³, blue surface indicate charge loss and yellow surfaces indicate charge accumulation), (e) optimized structures of the systems upon the interaction of K atom with an isolated OH group on GO basal planes, (f) the same as (e) but for the case of Li atom. The starting structures for (e) and (f) were as in **Figure 5** (b) and (d) (left). All the systems were calculated as charge neutral.

Even more striking differences in the behavior of isolated *versus* clustered OH groups can be seen in the results of DFT calculations (**Figure 6**). In this case, we performed calculations on charge-neutral systems and without implicit solvent. Hence, by adding one metal atom to the system, we simulate the addition of $M^+$ from the solution and electron through the external circuit. If Li or K atom is interacting with the OH groups cluster (**Figure 6**, a and b), no chemical change is observed. In other words, one *n*s electron from a metal atom is efficiently transferred to the GO sheet and stored in the basal plane. Although there is some charge depletion in the HO–C bond (**Figure 6**, c and d), the MOH unit is not detached. The situation is the opposite when M is brought in contact with an isolated OH group on the GO basal plane (**Figure 6**, d and e). In this case, the result is a spontaneous formation of MOH. These findings confirm that isolated OH groups over GO basal are more susceptible to reduction than clustered ones, agreeing with previous theoretical calculations [47]. Moreover, previous calculations have also shown that the interactions of K with clustered OH groups are energetically more favorable compared to the analogous interactions of Li with the same moieties.

Based on theoretical results, one could also understand why the apparent activation energy for electrochemical GO reduction increases with reduction potential to a certain point (**Figure 1**). This part of the $E_{act}$-electrode potential dependency could be correlated with the removal of easily reducible oxygen functional groups and the formation of reduced conductive domains with regenerated $sp^2$ hybridization, in line with experimental observations [18]. This hypothesis is also in line with the C-AFM measurements (**Figure 4**). Upon the progression of electrochemical GO reduction, hardly reducible oxygen groups are being removed, and breaking the structure of the O-moieties clusters makes their removal easier. For this reason, once the integrity of O-moieties clusters is compromised, the activation energy for reduction begins to decay (**Figure 1**).

## 4. Conclusions

Contrary to a thermally induced oxygen release, electrochemical reduction of GO in LiCl and KCl is a fast process with relatively low activation energy (below 30 kJ mol$^{-1}$), depending on the supporting electrolyte. The GO reduction occurs at lower potentials in KCl solution than in the LiCl solutions, which is observed not only by using cyclic voltammetry but also by simultaneous 2-point 4-point resistance measurements on thin GO films. While these measurements suggest that the reduced GO films become conductive when reduction reaches high degrees, C-AFM measurements show that reduction takes place locally but continuously and that conductive islands grow during the reduction of GO films. Once these islands coalesce, the lateral conductivity of the reduced GO films reaches its maximum, observed as a sharp decrease of the film resistance in s24 measurements. Our results also suggest that the evolution of gases during the GO film reduction ($H_2$, CO, $CO_2$) influences the contact between the reduced GO film and the substrate (in our case, Au electrode). This result is of high practical importance as a large contact resistance can cause significant energy losses in electrochemical energy conversion and storage applications of reduced GO films. Theoretical calculations show that the reduction of GO is more difficult if the oxygen functional groups are clustered over the GO basal plane. Hence, we propose that the GO reduction starts on isolated or low coordinated oxygen functional groups, which are easier to reduce. Then it progresses, causing a growth of the conductive islands. Presented results can help tailor reduced GO for capacitive and electrocatalytic applications. Such precise performance tuning can be enabled by exact control of the conductivity of the reduced GO films and of the amount of oxygen functional groups. This approach can be a way to maximize the performance of reduced GO in particular electrochemical applications.

**Acknowledgment**
I.A.P acknowledge the financial support provided by The Science Fund of the Republic of Serbia (PROMIS project RatioCAT) and Ministry of Education, Science and Technological Development of the Republic of Serbia (Contract No. 451-03-68/2020-14/200146); , I.A.P. and V.M.M. acknowledge DAAD Projekt "NANOCARBONS" (contract 57449323), S.J.G. and I.A.P. acknowledge the support provided by NATO Science for Peace and Security Programme, grant G5729; S.J.G acknowledges the financial support provided by the Federal Ministry of Education and Science of Bosnia and Herzegovina (Project: Funkcionalizovani grafenski materijali u elektrohemijskim sistemima za konverziju i skladištenje energije). B. V. acknowledges the funding provided by the Institute of Physics Belgrade through the grant of the Ministry of Education, Science, and Technological Development of the Republic of Serbia. The computations and data handling were enabled by resources provided by the Swedish National Infrastructure for Computing (SNIC) at the National Supercomputer Centre (NSC) at Linköping University, partially funded by the Swedish Research Council through grant agreement No. 2018-05973. N.V.S. acknowledges the support from Swedish Research Council (grant no 2019-05580).

**References**

[1] V. Dhinakaran, M. Lavanya, K. Vigneswari, M. Ravichandran, M.D. Vijayakumar, Review on exploration of graphene in diverse applications and its future horizon, in: Materials Today: Proceedings, Elsevier Ltd, 2020: pp. 824–828. https://doi.org/10.1016/j.matpr.2019.12.369.

[2] L. Wang, Z. Sofer, M. Pumera, Will Any Crap We Put into Graphene Increase Its Electrocatalytic Effect?, ACS Nano. 14 (2020) 21–25. https://doi.org/10.1021/acsnano.9b00184.

[3] W. Yuan, Y. Zhou, Y. Li, C. Li, H. Peng, J. Zhang, Z. Liu, L. Dai, G. Shi, The edge- and basal-plane-specific electrochemistry of a single-layer graphene sheet, Scientific Reports. 3 (2013). https://doi.org/10.1038/srep02248.

[4] A.S. Dobrota, I.A. Pašti, S. v. Mentus, N. v. Skorodumova, A general view on the reactivity of the oxygen-functionalized graphene basal plane, Physical Chemistry Chemical Physics. 18 (2016) 6580–6586. https://doi.org/10.1039/c5cp07612a.

[5] S. Georgitsopoulou, N.D. Stola, A. Bakandritsos, V. Georgakilas, Advancing the boundaries of the covalent functionalization of graphene oxide, Surfaces and Interfaces. 26 (2021). https://doi.org/10.1016/j.surfin.2021.101320.

[6] B.L. Dasari, J.M. Nouri, D. Brabazon, S. Naher, Graphene and derivatives – Synthesis techniques, properties and their energy applications, Energy. 140 (2017) 766–778. https://doi.org/10.1016/j.energy.2017.08.048.

[7] M.P. Araújo, O.S.G.P. Soares, A.J.S. Fernandes, M.F.R. Pereira, C. Freire, Tuning the surface chemistry of graphene flakes: new strategies for selective oxidation, RSC Advances. 7 (2017) 14290–14301. https://doi.org/10.1039/c6ra28868e.

[8] Y. Shao, J. Wang, M. Engelhard, C. Wang, Y. Lin, Facile and controllable electrochemical reduction of graphene oxide and its applications, Journal of Materials Chemistry. 20 (2010) 743–748. https://doi.org/10.1039/b917975e.

[9] S.Y. Toh, K.S. Loh, S.K. Kamarudin, W.R.W. Daud, Graphene production via electrochemical reduction of graphene oxide: Synthesis and characterisation, Chemical Engineering Journal. 251 (2014) 422–434. https://doi.org/10.1016/j.cej.2014.04.004.

[10] X. Zhang, D.C. Zhang, Y. Chen, X.Z. Sun, Y.W. Ma, Electrochemical reduction of graphene oxide films: Preparation, characterization and their electrochemical properties, Chinese Science Bulletin. 57 (2012) 3045–3050. https://doi.org/10.1007/s11434-012-5256-2.

[11] S.J. Gutić, D.K. Kozlica, F. Korać, D. Bajuk-Bogdanović, M. Mitrić, V.M. Mirsky, S. v. Mentus, I.A. Pašti, Electrochemical tuning of capacitive response of graphene oxide, Physical Chemistry Chemical Physics. 20 (2018) 22698–22709. https://doi.org/10.1039/c8cp03631d.

[12] D. Karačić, S. Korać, A.S. Dobrota, I.A. Pašti, N. v. Skorodumova, S.J. Gutić, When supporting electrolyte matters – Tuning capacitive response of graphene oxide via electrochemical reduction in alkali and alkaline earth metal chlorides, Electrochimica Acta. 297 (2019) 112–117. https://doi.org/10.1016/j.electacta.2018.11.173.

[13] J. Kauppila, P. Kunnas, P. Damlin, A. Viinikanoja, C. Kvarnström, Electrochemical reduction of graphene oxide films in aqueous and organic solutions, Electrochimica Acta. 89 (2013) 84–89. https://doi.org/10.1016/j.electacta.2012.10.153.

[14] O.C. Compton, S.T. Nguyen, Graphene oxide, highly reduced graphene oxide, and graphene: Versatile building blocks for carbon-based materials, Small. 6 (2010) 711–723. https://doi.org/10.1002/smll.200901934.

[15] W.J. Basirun, M. Sookhakian, S. Baradaran, M.R. Mahmoudian, M. Ebadi, Solid-phase electrochemical reduction of graphene oxide films in alkaline solution, Nanoscale Research Letters. 8 (2013) 397. https://doi.org/10.1186/1556-276X-8-397.

[16] J. Ping, Y. Wang, K. Fan, J. Wu, Y. Ying, Direct electrochemical reduction of graphene oxide on ionic liquid doped screen-printed electrode and its electrochemical biosensing application, Biosensors and Bioelectronics. 28 (2011) 204–209. https://doi.org/10.1016/j.bios.2011.07.018.

[17] S.J. Gutić, A.S. Dobrota, M. Leetmaa, N. v. Skorodumova, S. v. Mentus, I.A. Pašti, Improved catalysts for hydrogen evolution reaction in alkaline solutions through the electrochemical formation of nickel-reduced graphene oxide interface, Physical Chemistry Chemical Physics. 19 (2017) 13281–13293. https://doi.org/10.1039/c7cp01237c.

[18] J.S.D. Rodriguez, T. Ohigashi, C.-C. Lee, M.-H. Tsai, C.-C. Yang, C.-H. Wang, C. Chen, W.-F. Pong, H.-C. Chiu, C.-H. Chuang, Modulating chemical composition and work function of suspended reduced graphene oxide membranes through electrochemical reduction, Carbon. 185 (2021) 410–418. https://doi.org/10.1016/j.carbon.2021.09.015.

[19] Graphenea, Graphene oxide water dispersion, (2020). http://www.graphenea.com/collections/graphene-oxide/products/graphene-oxide-4-mg-ml-water-dispersion-1000-ml (accessed November 4, 2021).

[20] U. Lange, V.M. Mirsky, Separated analysis of bulk and contact resistance of conducting polymers: Comparison of simultaneous two- and four-point measurements with impedance measurements, Journal of Electroanalytical Chemistry. 622 (2008) 246–251. https://doi.org/10.1016/j.jelechem.2008.06.013.

[21] V. Kulikov, V.M. Mirsky, T.L. Delaney, D. Donoval, A.W. Koch, O.S. Wolfbeis, High-throughput analysis of bulk and contact conductance of polymer layers on electrodes, Measurement Science and Technology. 16 (2005) 95–99. https://doi.org/10.1088/0957-0233/16/1/013.

[22] Q. Hao, V. Kulikov, V.M. Mirsky, Investigation of contact and bulk resistance of conducting polymers by simultaneous two- and four-point technique, Sensors and Actuators, B: Chemical. 94 (2003) 352–357. https://doi.org/10.1016/S0925-4005(03)00456-8.

[23] G. Kresse, J. Hafner, *Ab initio* molecular dynamics for liquid metals, Physical Review B. 47 (1993) 558–561. https://doi.org/10.1103/PhysRevB.47.558.

[24] G. Kresse, J. Furthmüller, Efficiency of ab-initio total energy calculations for metals and semiconductors using a plane-wave basis set, Computational Materials Science. 6 (1996) 15–50. https://doi.org/10.1016/0927-0256(96)00008-0.

[25] G. Kresse, J. Furthmüller, Efficient iterative schemes for *ab initio* total-energy calculations using a plane-wave basis set, Physical Review B. 54 (1996) 11169–11186. https://doi.org/10.1103/PhysRevB.54.11169.

[26] J.P. Perdew, K. Burke, M. Ernzerhof, Generalized Gradient Approximation Made Simple, Physical Review Letters. 77 (1996) 3865–3868. https://doi.org/10.1103/PhysRevLett.77.3865.

[27] P.E. Blöchl, Projector augmented-wave method, Physical Review B. 50 (1994) 17953–17979. https://doi.org/10.1103/PhysRevB.50.17953.

[28] J.J.P. Stewart, Stewart computational chemistry - MOPAC, (2016). http://openmopac.net/ (accessed November 4, 2021).

[29] J.J.P. Stewart, Optimization of parameters for semiempirical methods VI: more modifications to the NDDO approximations and re-optimization of parameters, Journal of Molecular Modeling. 19 (2013) 1–32. https://doi.org/10.1007/s00894-012-1667-x.

[30] A. Klamt, G. Schüürmann, COSMO: A new approach to dielectric screening in solvents with explicit expressions for the screening energy and its gradient, Journal of the Chemical Society, Perkin Transactions 2. (1993) 799–805. https://doi.org/10.1039/P29930000799.

[31] K. Momma, F. Izumi, VESTA 3 for three-dimensional visualization of crystal, volumetric and morphology data, Journal of Applied Crystallography. 44 (2011) 1272–1276. https://doi.org/10.1107/S0021889811038970.

[32] Jmol, Jmol: an open-source Java viewer for chemical structures in 3D., (2010). http://www.jmol.org/ (accessed November 4, 2021).

[33]  I. Jung, D.A. Field, N.J. Clark, Y. Zhu, D. Yang, R.D. Piner, S. Stankovich, D.A. Dikin, H. Geisler, C.A. Ventrice, R.S. Ruoff, Reduction kinetics of graphene oxide determined by electrical transport measurements and temperature programmed desorption, Journal of Physical Chemistry C. 113 (2009) 18480–18486. https://doi.org/10.1021/jp904396j.

[34]  O.M. Slobodian, P.M. Lytvyn, A.S. Nikolenko, V.M. Naseka, O.Yu. Khyzhun, A. v. Vasin, S. v. Sevostianov, A.N. Nazarov, Low-Temperature Reduction of Graphene Oxide: Electrical Conductance and Scanning Kelvin Probe Force Microscopy, Nanoscale Research Letters. 13 (2018) 139. https://doi.org/10.1186/s11671-018-2536-z.

[35]  K. Yin, H. Li, Y. Xia, H. Bi, J. Sun, Z. Liu, L. Sun, Thermodynamic and Kinetic Analysis of Lowtemperature Thermal Reduction of Graphene Oxide, Nano-Micro Letters. 3 (2011) 51–55. https://doi.org/10.1007/BF03353652.

[36]  Y. Qiu, F. Collin, R.H. Hurt, I. Külaots, Thermochemistry and kinetics of graphite oxide exothermic decomposition for safety in large-scale storage and processing, Carbon. 96 (2016) 20–28. https://doi.org/10.1016/j.carbon.2015.09.040.

[37]  S. Xue, B. Garlyyev, S. Watzele, Y. Liang, J. Fichtner, M.D. Pohl, A.S. Bandarenka, Influence of Alkali Metal Cations on the Hydrogen Evolution Reaction Activity of Pt, Ir, Au, and Ag Electrodes in Alkaline Electrolytes, ChemElectroChem. 5 (2018) 2326–2329. https://doi.org/10.1002/celc.201800690.

[38]  A. Viinikanoja, Z. Wang, J. Kauppila, C. Kvarnström, Electrochemical reduction of graphene oxide and its in situ spectroelectrochemical characterization, Physical Chemistry Chemical Physics. 14 (2012) 14003–14009. https://doi.org/10.1039/c2cp42253k.

[39]  A.C. Ferrari, J. Robertson, Interpretation of Raman spectra of disordered and amorphous carbon, Physical Review B. 61 (2000) 14095–14107. https://doi.org/10.1103/PhysRevB.61.14095.

[40]  R.R. Nair, W. Ren, R. Jalil, I. Riaz, V.G. Kravets, L. Britnell, P. Blake, F. Schedin, A.S. Mayorov, S. Yuan, M.I. Katsnelson, H.M. Cheng, W. Strupinski, L.G. Bulusheva, A. v. Okotrub, I. v. Grigorieva, A.N. Grigorenko, K.S. Novoselov, A.K. Geim, Fluorographene: A two-dimensional counterpart of Teflon, Small. 6 (2010) 2877–2884. https://doi.org/10.1002/smll.201001555.

[41]  A. Eckmann, A. Felten, A. Mishchenko, L. Britnell, R. Krupke, K.S. Novoselov, C. Casiraghi, Probing the nature of defects in graphene by Raman spectroscopy, Nano Letters. 12 (2012) 3925–3930. https://doi.org/10.1021/nl300901a.

[42]  J.A. Quezada Renteria, C. Ruiz-Garcia, T. Sauvage, L.F. Chazaro-Ruiz, J.R. Rangel-Mendez, C.O. Ania, Photochemical and electrochemical reduction of graphene oxide thin films: Tuning the nature of surface defects, Physical Chemistry Chemical Physics. 22 (2020) 20732–20743. https://doi.org/10.1039/d0cp02053b.

[43]   A. Lerf, H. He, M. Forster, J. Klinowski, Structure of Graphite Oxide Revisited, The Journal of Physical Chemistry B. 102 (1998) 4477–4482. https://doi.org/10.1021/jp9731821.

[44]   X. Gao, J. Jang, S. Nagase, Hydrazine and thermal reduction of graphene oxide: Reaction mechanisms, product structures, and reaction design, Journal of Physical Chemistry C. 114 (2010) 832–842. https://doi.org/10.1021/jp909284g.

[45]   S. Stankovich, D.A. Dikin, R.D. Piner, K.A. Kohlhaas, A. Kleinhammes, Y. Jia, Y. Wu, S.B.T. Nguyen, R.S. Ruoff, Synthesis of graphene-based nanosheets via chemical reduction of exfoliated graphite oxide, Carbon. 45 (2007) 1558–1565. https://doi.org/10.1016/j.carbon.2007.02.034.

[46]   D.W. Boukhvalov, M.I. Katsnelson, A.I. Lichtenstein, Hydrogen on graphene: Electronic structure, total energy, structural distortions and magnetism from first-principles calculations, Physical Review B. 77 (2008) 035427. https://doi.org/10.1103/PhysRevB.77.035427.

[47]   A.S. Dobrota, S. Gutić, A. Kalijadis, M. Baljozović, S. v. Mentus, N. v. Skorodumova, I.A. Pašti, Stabilization of alkali metal ions interaction with OH-functionalized graphene: Via clustering of OH groups-implications in charge storage applications, RSC Advances. 6 (2016) 57910–57919. https://doi.org/10.1039/c6ra13509a.